# Personalization of Hearing Aid Compression by Human-In-Loop Deep Reinforcement Learning

N. Alamdari[1], *Student Member, IEEE*, E. Lobarinas[2], and N. Kehtarnavaz[1], *Fellow, IEEE*

[1]Electrical and Computer Engineering Department, University of Texas at Dallas, TX 75080 USA
[2]Callier Center for Communication Disorders, University of Texas at Dallas, TX 75080 USA
Corresponding author: N. Alamdari (e-mail: alamdari@utdallas.edu).

*Abstract*—**Existing prescriptive compression strategies used in hearing aid fitting are designed based on gain averages from a group of users which are not necessarily optimal for a specific user. Nearly half of hearing aid users prefer settings that differ from the commonly prescribed settings. This paper presents a human-in-loop deep reinforcement learning approach that personalizes hearing aid compression to achieve improved hearing perception. The developed approach is designed to learn a specific user's hearing preferences in order to optimize compression based on the user's feedbacks. Both simulation and subject testing results are reported which demonstrate the effectiveness of the developed personalized compression.**

## I. INTRODUCTION

In hearing impaired individuals, the relative intensity difference between barely audible and uncomfortably loud sound becomes smaller. Thus, in order to achieve optimal audibility, sound must be calibrated to occupy a smaller range of sound pressure levels (SPLs). This dynamic range adjustment is achieved through the process of compression [1]. Compression in a reduced dynamic range is the key function of modern hearing aids. This process involves squeezing or fitting sound into the audible range of a hearing aid user. In hearing aid fitting, so-called compression curves are set up by adjusting gains across a number of frequency bands based on a user's audiometric profile. The two most widely used hearing aid prescriptions are NAL-NL2 [2] and DSL-v5 [3]. These prescriptions correspond to gain tables across a number of frequency bands for three sound levels, soft, moderate, and loud.

It has been reported that up to half of individuals using fitted hearing aids preferred amplification or compression settings different than the prescription provided [4-8]. Considering that suprathreshold hearing perception varies from person to person and that acoustic environments encountered vary from person to person, several papers in the literature have examined self-adjustment or self-tuning of hearing aid fitting relative to the one-size-fits-all prescriptive fitting [9-16]. In [16], it was reported that hearing aid users favored gain settings that were different from the NAL prescription settings both in quiet and in noise. Self-adjustments carried out on a custom hardware such as the one recently reported in [17] have also demonstrated improvements in hearing perception that can be gained over prescriptive fitting.

In [18-20], a machine learning approach for self-adjustment or self-tuning of compression was presented. In these papers, a Gaussian regression model was used to achieve personalized compression by estimating its parameters from training data. User preferences were obtained via hearing assessments by listening to music clips in [18,19]. Although the results reported show the benefits of personalization, differences in preferences between music clips and conversation in noisy environments (e.g., in babble noise) were not addressed. Understanding speech in the presence of bothersome background noise is expressed as a major challenge by hearing aid users [21]. Furthermore, in [18-20], only twenty preference iterations were done for modeling the hearing preference of a user. In actual audio environments, this many iteration would be inadequate for modeling various non-linearities associated with hearing perception.

To address the non-linearities of hearing perception, a human-in-loop interactive machine learning approach based on deep reinforcement learning (DRL) [22] is developed in this paper. In our approach, the user is placed in the learning loop. A DRL network is designed to receive preference feedback from the user. As a result, it becomes possible to deal with various non-linearities of human hearing perception. A combination of a Convolutional Neural Network (CNN) [23] and a Bidirectional Long Short-Term Memory Recurrent Neural Network [24] or CNN-BiLSTM is used to model a user's preferences in those audio environments that are of interest to the user. As discussed in [25], user feedback is affected by biases. Thus, rather than absolute feedback, pairwise or relative hearing assessments are deemed more suitable [26]. Hence, in our approach, a user is subjected to a series of compressed audios to express his/her preferences towards training the model via the reward/punishment mechanism of reinforcement learning; an approach that has been successfully applied to gaming [27] and robotics [28]. The developed DRL approach provides personalized compression that can be utilized in the field for hearing aid compression studies.

To describe our approach in detail, the remainder of this paper is organized as follows. Section II covers the developed approach to personalize hearing aid compression or fitting via human-in-loop deep reinforcement learning as well as a protocol to perform human preference assessment. The experimental results and their discussion are then stated in section III followed by the conclusion in section IV.

## II. PERSONALIZED COMPRESSION APPROACH

To set the stage for the developed personalized compression, the conventional reinforcement learning (RL) is first briefly described. In a reinforcement learning system, an agent and an environment interact over a series of steps. At each time step $t$, the agent receives an observation or state $s_t \epsilon S$ from the envoriment and sends an action $a_t \epsilon A$ back to the environment with $S$ and $A$ denoting the state and action set, respectively. In a conventional RL system, based on a given action, the environment generates the next state together with a reward $r_t \epsilon R$ with $R$ denoting the reward set, and the goal is to maximize reward over time. Fig. 1(a) shows a block diagram of a conventional RL system.

The success of RL heavily depends on setting up an effective reward function. Many real-world problems are complex and it is often difficult to formulate an effective reward function. Inverse reinforcement learning (IRL) [29] can be used to design a reward function, which can then get deployed to train the agent using (deep) reinforcement learning. In order to build an effective reward function, human feedback can be used to evaluate the behavior of the agent [30, 31]. In our case, in order to model and learn hearing preferences via deep reinforcement learning, the listener's preferences are used.

Obtaining rewards in a direct manner, based on user feedback, is labor intensive and makes the training process impractical because thousands of iterations and user feedbacks would be needed. In order to decrease the number of user feedbacks and thus enable a practical deployment of the personalized compression, first a reward function is considered to model hearing preferences of a user in an asynchronous manner. This is achieved by carrying out comparison between instances of two different compressed audios. Then, an agent is trained to maximize reward. Fig. 1(b) shows a block diagram of this approach. Unlike the conventional RL in which reward is computed by the environment, here reward is computed based on user preferences.

### A. PERSONALIZED FITTING PROTOCOL

Compression or hearing aid fitting is normally performed via software tools that are provided by hearing aid manufacturers. These software tools are used to set gains across a number of frequency bands using established prescriptions based on group averages or with manufacturers adding their own variations. A user's audiogram and the prescription gains are used to set the target gains for that user. Across each frequency band, a different compression curve is used to generate multiband dynamic range compression (DRC). In this paper, the DSL-v5 by Hand prescription in [32] is considered to serve as the reference compression. In other words, the gains in the DSL-v5 tables are used to compute the reference DRC parameters consisting of compression ratio (change in gain) and compression threshold (sound level at which compression is applied). The process of personalization involves modification of the gains specified by DSL-v5 or any other generic prescription based on user preferences.

The steps taken here to achieve personalized compression for a specific user are: (i) measuring the audiogram of the user, (ii) defining the compression gains of the user using a fitting software, DSL-v5 here, (iii) initializing the human centered-DRL system with the compression ratios obtained in (ii) as the starting point, (iv) adjusting the compression ratios by going through the training process of the human centered-DRL system (an illustration of the gain change ranges for the agent action in the DRL system is depicted in Fig. 2, and (v) comparing the performance of the personalized compression with the prescriptive or reference compression.

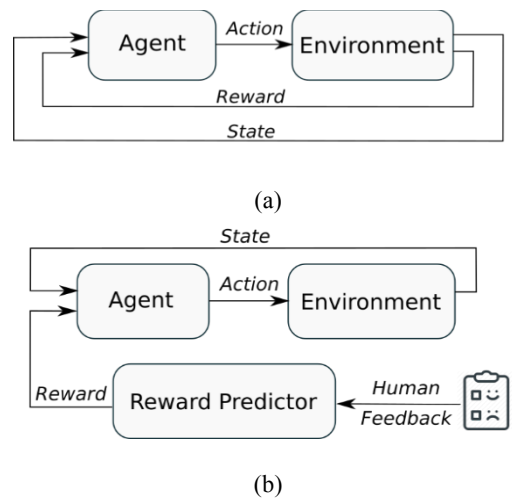

(a)

(b)

**FIGURE 1.** (a) Block diagram of a conventional reinforcement learning system. (b) Deep reinforcement learning with user feedback in which reward is obtained based on user preferences.

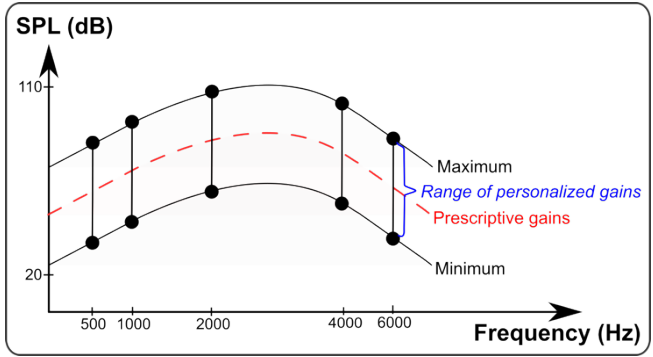

**FIGURE 2.** Illustration of gain ranges across different frequency bands.

In the personalized system, an agent that is interacting sequentially with the environment over a number of time steps is considered similar to the one in [30]. At each time step, the agent receives a new state (observation) from the environment and performs an action. Over an episode, the agent performs interaction for a number of time steps. In typical RL settings, a reward at each time step is fed into the agent as well. However, here rather than a predefined reward, a reward that is modeled based on a user's preferences is considered. In other words, by putting human feedback in the learning loop, it is attempted to optimize the agent learning and thus the user hearing perception. The personalization protocol and block diagram of the developed DRL-based personalized hearing aid compression system is shown in Table 1 and Fig. 3, respectively. Each block in Fig. 3 is described in the subsections that follow.

### B. ENVIRONMENT

In the personalized system, the environment consists of three components: *audio segment creation*, *compression ratio update*, and *agent state transition function*. At each policy time step, a noisy speech audio signal is down-sampled from 48 kHz to 16 kHz to lower the computational burden. Noisy speech audio signals are generated by distorting the widely used public domain IEEE speech dataset [33] with babble restaurant noise (provided on YouTube) and SNR of about 0 dB. The IEEE dataset consists of 3600 speech audio files by 20 speakers (10 females and 10 males) in which each file is about 2 seconds long. The speakers are from two American English regions of the Pacific Northwest (PN) and the Northern Cities (NC) reading the IEEE "Harvard" sentences. A total of 3600 noisy speech audio signals are thus generated and at each time step, a randomly selected audio signal is used for preference training.

Once a new action $a_{t+1}$ is received from the agent, CR (compression ratio) in the frequency bands are updated based on the action $a_{t+1}$. Depending on the possible number of scales ($\beta$) considered in adjusting CR in the frequency bands, a set of actions are created by permutations and the action set $A$ is given by

$$A = \prod_i^\beta A_i \qquad (1)$$

$$CR_{new}(f) = CR_{DSL-v5}(f) \cdot CR_{adj}(f) \qquad (2)$$

where $CR_{adj}(f)$, $CR_{DSL-v5}(f)$, and $CR_{new}(f)$, respectively, stand for the compression ratio adjustment, the compression ratio computed from the DSL-v5 prescription, and the new compression ratio in the $f^{th}$ frequency band. Permutations are defined by a dictionary in which each action is mapped to a set of compression ratio adjustments in the frequency bands.

TABLE I
DEVELOPED PERSONALIZED COMPRESSION

| Personalized fitting protocol |
|---|
| 1. Measuring audiogram of the user |
| 2. Defining the compression gains of the user using a prescriptive fitting software (e.g., DSL-v5 in [3]); computing compression ratios in a number of frequency bands |
| 3. Initializing the human-centered DRL system with the above compression ratios |
| 4. Running the policy in the environment and storing a set of compression ratio adjustments that are resulted from randomly generated agent's actions; |
| 5. Generating pairs of compressed audio signals with compression ratios in step 4 |
| 6. Asking the user to label each pair and add audio pairs and their labels to a buffer |
| 7. Training the preference (reward) predictor using the buffer |
| 8. Training the RL policy, based on the observation received from the environment with the reward from the trained reward predictor. |
| 9. **For** *M* iterations **do**: |
| 10.     Training the policy in the environment for $N_{steps}$ with the reward from the reward predictor |
| 11.     Selecting compressed audio pairs resulted from given actions |
| 12.     Asking the user to declare preferences and adding preferences to a buffer |
| 13.     Fine-tuning the reward model for *k* batches from the above buffer |
| 14. **End for** |
| 15. Hearing assessment to compare personalized compression ratios with those specified by DSL-v5 prescriptive compression |

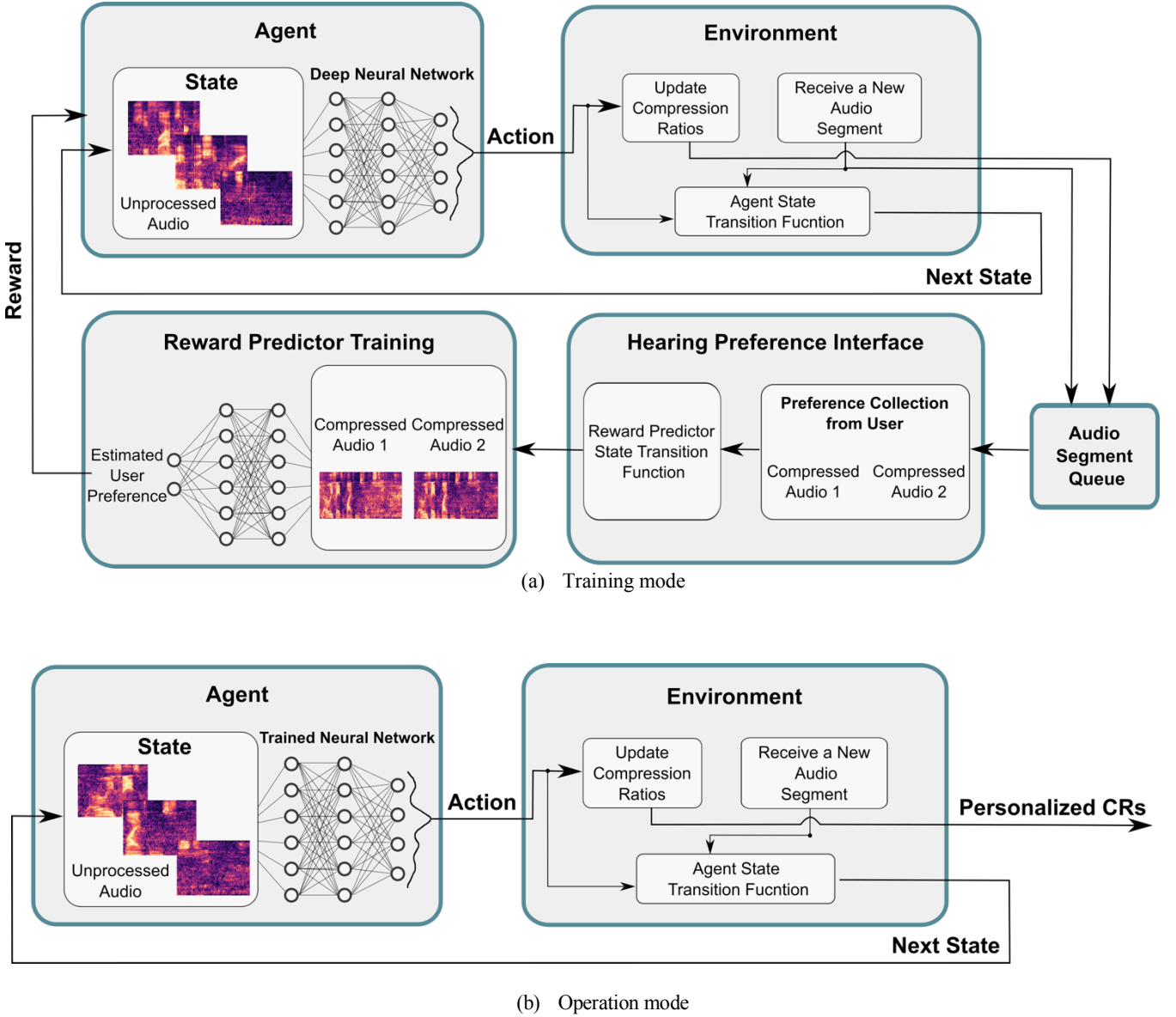

FIGURE 3. Developed personalized compression DRL system: (a) training mode and (b) operation mode.

In *agent state transition function*, a new audio signal is compressed by the updated CRs from the previous iteration using the dynamic range compression whose details are described in our previous publication [32]. The compressed noisy speech signals are placed into 20ms frames with 10ms overlap. Log Mel-spectrogram features are extracted from each frame using a bank of 80 Mel filters, which are then stacked to form a log Mel-spectrogram image of size 80*80. The stacking of three such adjacent images is considered to be one observation. In other words, each compressed audio signal is mapped to a stack of 3 log Mel-spectrogram images of size 80*80. Note that in contrast to the conventional RL, here the end-of-episode sign and reward values from the environment are not shared with the agent to make the agent training one uninterrupted episode.

Next, unprocessed audio signals and updated CRs are added to a buffer called *audio segment queue* as depicted in Fig. 3(a). The audio segment queue σ shown in this figure denotes a set or collection of audio signals $U$ and compression ratios $CR$ computed from a number of $k$ actions, that is

$$\sigma = ((u_0, cr_0), (u_1, cr_1), \ldots, (u_{k-1}, cr_{k-1})) \epsilon (U, CR)^k \quad (3)$$

## C. HUMAN PREFERENCE INTERFACE

In order to use human input in the learning loop, a hearing preference interface is created to collect the user's hearing feedbacks from a group of comparisons of audio signal pairs that are compressed with two different sets of compression ratios. The goal of this user interaction is to learn the non-linearities associated with the user's preferences or reward function.

In *hearing preference interface* shown in Fig. 3, two pairs from the queue $\sigma$ are selected at each time step. Then, a corresponding pair of compressed audio signals $(c^1, c^2)$ is computed which is used for the comparison. The user is given 4 options to indicate his/her preference: (1) $\mu = [1,0]$ if $c^1$ is preferable, (2) $\mu = [0,1]$ if $c^2$ is preferable, (3) $\mu = [0.5, 0.5]$ if both compressed audio signals are equally preferred, and (4) neither compressed audio signals are desired. Hearing preferences are collected over a series of compressed audio signal pairs and are stored in a dataset $D$ of triplets $(c^1, c^2, \mu)$, where $\mu$ denotes the feedback label, and $c^1$ and $c^2$ are the two compressed audio signals created by applying two different sets of CRs to the same noisy speech signal. Note that for option (4), the comparison is excluded from the dataset $D$.

In *reward predictor state transition function*, a batch of data from $D$ is used to train the reward predictor model to improve the agent policy. Similar to the agent's state transition function, each compressed audio signal is framed into 20ms frames with 50% overlap. Log Mel-spectrogram features of each frame are computed and considered to be one observation.

**Data augmentation** - The performance of the human hearing preference estimator depends on both the number of feedbacks acquired from the user and the model structure. A data augmentation is thus performed to address the limited size of training data that is available to the reward estimator in practice. This data augmentation consists of creating realistic samples by substituting features of audio signal 1 with features of audio signal 2. Their corresponding preference label is switched accordingly. The goal of the data augmentation here is to enhance the generalization capability of the preference (reward) predictor.

In addition to increasing the size of the training data, it is made sure that the training data does not suffer from unbalanced labels. Unbalanced labels can cause the model not to learn the learning preferences due to: (1) the network model not getting optimized for the unbalanced label in the original dataset, and (2) the accuracy of a validation or test set drops as it is challenging to have a complete representation with few observations.

## D. PREFERENCE / REWARD PREDICTOR

In the reward predictor block shown in the Fig. 3, the parameters reflecting the reward are obtained via a combination of a Convolutional Neural Networks (CNN) and a Bidirectional Long Short-Term Memory (CNN-BiLSTM) in a supervised manner. Log Mel-spectrogram features of compressed audio pairs constitute the input of the network and user feedbacks constitute the output of the network. The reward or hearing preference predictor provides a reward prediction $\hat{r}$ and produces the probability associated with preferring a compressed audio signal $c^1$ over another compressed audio signal $c^2$. For the prediction $\hat{r}$, the following cross-entropy loss function between the predicted reward and the actual user feedback is minimized:

$$loss(\hat{r}) = -\sum_{(c^1, c^2, \mu) \in D}(\mu(1) \log \hat{P}[c^1 > c^2] +$$
$$\mu(2) \log \hat{P}[c^1 < c^2]) \qquad (4)$$

Learning preferences and predicting reward from comparison pairs poses an implementation difficulty as a comparison pair does not provide a numeric feedback. To estimate the agent reward, one needs to estimate it from an intermediate model or network. The network structure of the developed hearing preference predictor is depicted in Fig. 4. Fig. 4(b) shows the overall structure of the network used for training the reward predictor based on the dataset of pair comparisons. Batch normalization [34] is applied to the convolutional layers using a decay rate of 0.90 together with a dropout with alpha = 0.5. During the training phase, the model is trained on a batch size of 64, and optimized using the Adam algorithm [35]. Furthermore, during the training phase, early stopping and adaptive learning rate are applied to avoid overfitting. Once the reward predictor is trained, the intermediate model or shared network as depicted in Fig. 4 (c) is used for agent training. Due to the fact that DRL is sensitive to the reward scale, a sigmoid layer is added at the end of the shared reward predictor model to bring the predicted reward between 0 and 1, see Fig. 4 (c).

## E. RL Agent

The training for a RL policy $\pi$ is carried out based on the Bellman equation [36] in which at each time step $t$, the RL policy provides an action $a_t$ for a given state $s_t$ as expressed below

$$\pi: S \to A \qquad (5)$$
$$a_t = \pi(s_t)$$

Action $a_t$ influences the future state of the agent. The success of RL in learning the policy is reflected in the reward and the goal of RL is to maximize the overall reward. The parameters of the policy can get updated based on deep Q-learning [36] to maximize the overall estimated reward $\hat{r}$. Here, Q-value in Q-learning is optimized by a convolutional neural network (CNN). Q-value at a time step $j$ is computed as follows:

$$y_j \leftarrow r(s_j, a_j) + \gamma \max_{a'} Q_\phi(s'_j, a'_j) \qquad (6)$$

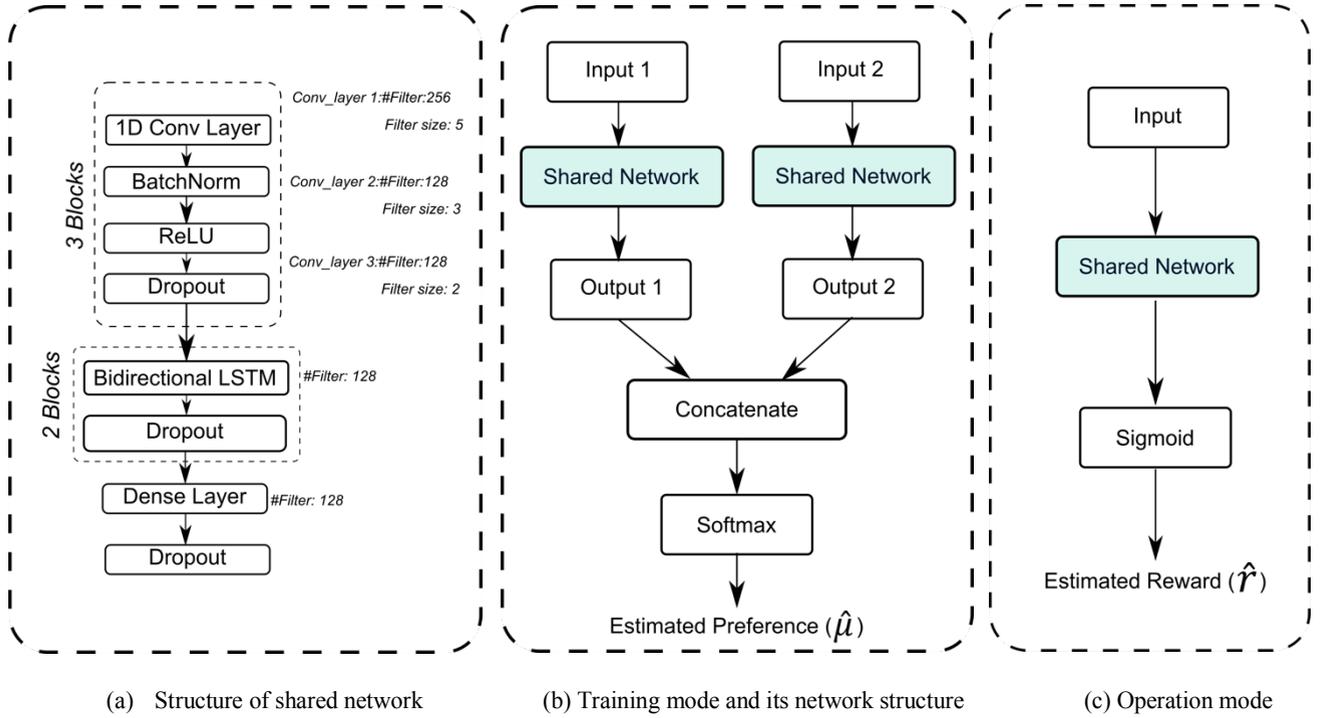

(a) Structure of shared network  (b) Training mode and its network structure  (c) Operation mode

**FIGURE 4.** Network structure of the reward predictor.

where $r(s, a)$ denotes the reward of a state and an action, and $\gamma$ is a discount factor. A Q-value is basically a prediction of the future reward which allows selecting a next action for a given state. To convert the output values of the CNN into action probabilities, the so-called one-hot representation is utilized. As noted below, the action with the highest probability is then selected

$$a_j = \arg max_a Q_\phi(s_j, a_j) \quad (7)$$

The loss function in this CNN-based Q-learning is the mean-square-error (MSE) and the optimization is done by using the Adam algorithm, resulting in the CNN weights to get updated as follows:

$$\phi \leftarrow \arg min_\phi \|Q_\phi(s_j, a_j) - y_j\|^2 \quad (8)$$

It is important to note that in contrast to supervised learning in which targets are fixed before training, here targets of the CNN-based agent depend on the network's weights that get updated gradually.

Before starting to train the RL agent and in order to reduce the chance of training a bad policy based on an untrained reward predictor, the reward predictor is trained with the dataset $D$ of user preferences (mentioned earlier in subsection D). This means that the training of the reward predictor is performed asynchronously with respect to the DRL agent. For example, 200 comparison pairs can be conducted by the user at the beginning of the DRL training. Then, querying of the user feedback can be done every $M$ time steps (see Table I).

For the CNN $Q_\phi(s_j, a_j)$ model, a similar configuration used in the Atari experiment in [31] is utilized here. That is 80*80*3 stacked unprocessed audio frames are used as the input to the policy consisting of 3 convolutional layers having 32, 64, and 128 filters, respectively, with rectified linear unit (ReLU) activation and alpha = 0.01. Then, the flattened output of the last convolution layer is concatenated with the compression ratio adjustments ($CR_{adj}$) of the previous time step. This is followed by two fully-connected layers of size 256 with ReLU activation, and a fully-connected layer with a size equal to the action space size. A fraction of the dataset is used as validation data to avoid overfitting. The agent is trained for 300 episodes, each containing 20 agent time steps. The reward model is fixed during the training. The value and description of parameters associated with the agent training are summarized in Table II.

In this approach, non-numerical feedback rather than absolute feedback is obtained from a user. The goal is to learn a policy that is most consistent with the user's preferences. As a result, personalization emulates the intention of the user and finds a policy that is ideally consistent with it. The above learning can be viewed as an active learning approach for achieving personalized compression. This approach has the advantage of being able to get trained in an online manner, thus allowing its utilization in the field or in real-world audio environments.

TABLE II
PARAMETERS ASSOCIATED WITH AGENT TRAINING

| Parameter | Value | Description |
|---|---|---|
| $N_{Episode}$ | 300 | Number of episodes for training |
| $N_{Steps}$ | 20 | Number of steps in an episode |
| Training frequency | 20 | Number of steps to train agent |
| Batch size | 50 | Number of training observations used in one iteration |
| $\gamma$ | 0.99 | Discount factor in updating Q-learning |
| No-op | 30 | Number of time steps before starting to train the agent |

## III. EXPERIMENTAL RESULTS

Two sets of experiments were conducted to examine the performance of the developed personalized DRL compression. The first set included simulations of the human-in-loop deep reinforcement learning. In the second set of experiments, five adult human participants with bilateral, mild to moderate hearing loss were tested. All human subject testing was performed under an approved IRB (Institutional Review Board) protocol at the University of Texas at Dallas. In the two subsections that follow, these two sets of experiments are described. The purpose of the simulation experiments was to show the capability of the developed personalized approach to learn hearing preferences towards generating compression gains that best matched a specific setting or user. In addition, the results of the personalized compression for five participants with hearing loss are reported.

### A. Simulation Experiments

In the first set of experiments, five hearing preference scenarios were simulated and the outcomes were examined to see the learning capability of the developed personalized DRL compression. Naturally, as described earlier, when the size of the action space is increased by increasing the number of frequency bands or number of scales $\beta$, the personalization demands more iterations posing difficulty for its actual deployment. To illustrate this point, five simulated hearing aid users were considered. For simulated hearing aid users 1, 2, and 3, five frequency bands were considered and for simulated hearing aid users 4 and 5 only two and three frequency bands (first, third, and fifth bands), were considered, respectively.

For all the simulated hearing aid users, the same audiogram or the same prescription gains from [32] were used. Then, the DSL-v5 prescriptive gains expressed in nine frequency bands were mapped to five bands to reduce the computational complexity. These five frequency bands were: [0-500] Hz, [500-1000] Hz, [1.0-2.0] kHz, [2.0-4.0] kHz, and [4.0-6.0] kHz. The compression ratios (gain changes) were computed from the gains. The attack time (time it takes to respond to higher sound levels) and the release time (time it takes to respond to lower sound levels) were set to the typical values of 0.01s and 1.0s, respectively. Basically, the attack and release time regulate the reaction pace of compression. Moderate and loud compression thresholds were also set to 60dB and 80dB to be consistent with the prescriptive hearing aid compression fittings. As a result, the action space of the first three simulated users became as depicted in Fig. 5(a), exhibiting 32 possible compression ratio adjustments in five frequency bands. Fig. 5(b) depicts the action space of a simpler case by changing the compression ratios in only the first and the fifth frequency bands. The action space for the case of three frequency bands (first, third, and fifth) is depicted in Fig. 5(c).

Overall, 200 hearing preferences over audio pairs were considered for each simulated user. As illustrated in Fig. 6, when the action space became larger, more user feedbacks were required. This figure not only shows the complexity of hearing preferences in larger action spaces, but also differences in preferences between the users 1, 2, and 3. The simulated user 2 is an example of the situation when an actual user does not give proper feedback preferences when listening to audio pairs of two sets of compression ratios. This led to failure in learning preferences during the reward predictor training which consequently led to failure in the agent learning or learning the best settings for that user. The simulated user 3 is an example of the situation when an actual user is very strict about some settings and has neutral preferences over the other settings. This led to having more neutral preferences and therefore the training dataset became highly unbalanced.

The learning loss value of hearing preferences with respect to training epochs in different scenarios is displayed in Fig 7. By comparing Fig. 7(a) with Fig. 7(d), it can be seen that the learning process in the reward predictor training became more challenging as the action space became larger. In addition, Fig. 7(b) illustrates a poor learning of the reward predictor due to inadequate feedbacks from the user.

The mean of the normalized reward and the mean of the Q-values (target outputs) across the agent training episodes for each simulated user are displayed in Figs. 8 and 9, respectively. From these figures, it can be seen that both the mean reward and the mean Q-value exhibited an increasing trend, indicating that the personalized compression was gradually learning the policy that was ideally consistent with the users' hearing preferences. As mentioned earlier, the success of RL is heavily dependent on the performance of the reward predictor. That is why, although an increasing trend for user 2 is exhibited in Fig. 8(b) and Fig. 9(b), the personalization was not effective due to the poor training of the reward predictor.

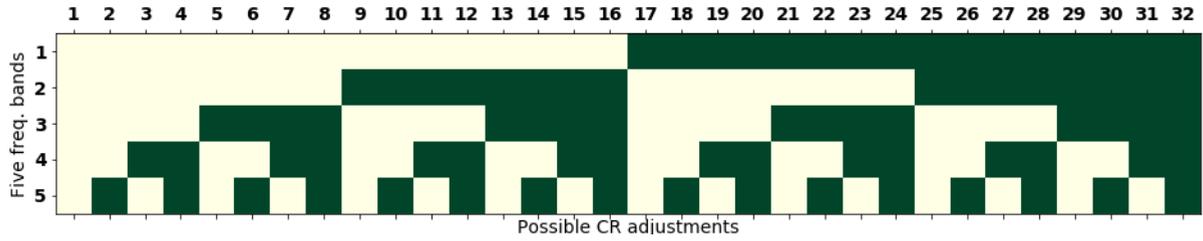

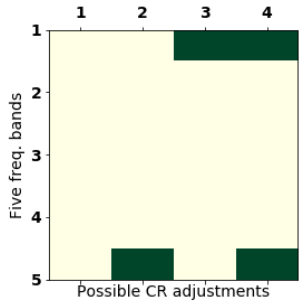

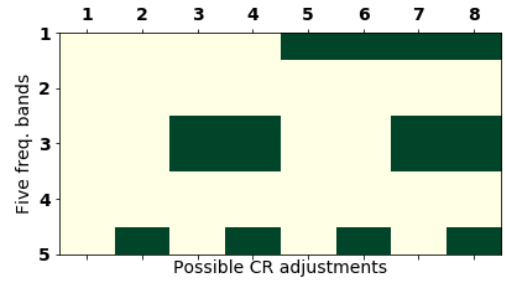

(a) Adjustments in five frequency bands that is mapped to 32 actions

(b) Adjustments in the first and the fifth frequency bands

(c) Adjustments in the first, the third, and the fifth frequency bands

**FIGURE 5. Action space corresponding to (a) simulated users 1, 2, and 3, (b) simulated user 4, and (c) simulated user 5 with β = 2 ($CR_{adj} = 1\ or\ 4$) and five frequency bands; light color indicates $CR_{adj} = 1$ and dark color indicates $CR_{adj} = 4$.**

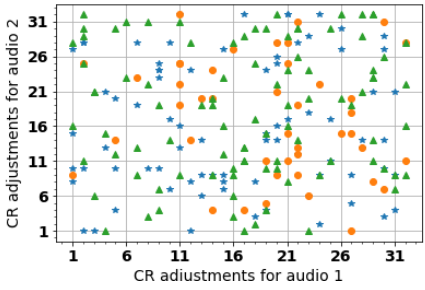
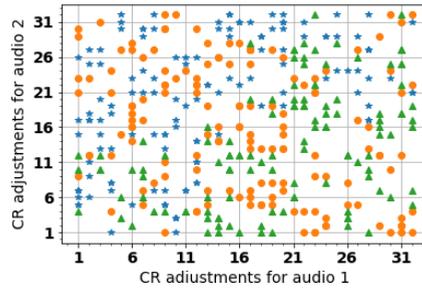
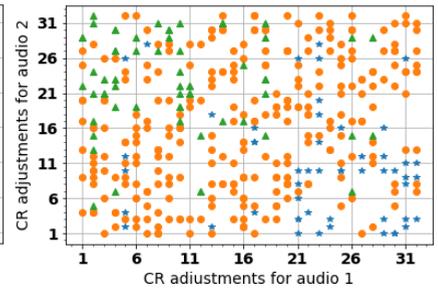

(a) User 1      (b) User 2      (c) User 3

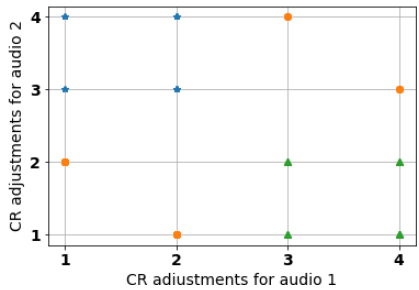
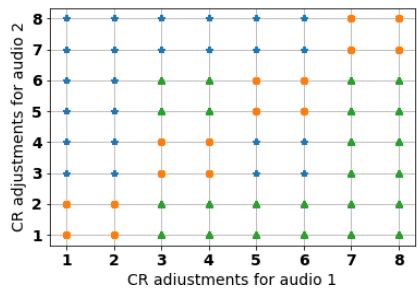

(d) User 4      (e) User 5

**FIGURE 6. Preference space collected from (a) simulated user 1, (b) simulated user 2, (c) simulated user 3, (d) simulated user 4, and (e) from simulated user 5.**

## B. Subject Testing Experiments

In addition to the above simulations, actual human subject testing was performed according to the IRB protocol described earlier with one modification. Due to the Covid-19 pandemic, the original IRB was modified to allow participant testing to be conducted online instead of in a soundbooth. For the subject testing experiments, "virtual" visits were conducted using the Skype video conference utility. Secure links were emailed to the participants to access online experimental sessions. Initially, the participants obtained their audiograms using the web-based hearing test at https://hearingtest.online/.

As per Table I, for training the developed deep reinforcement learning personalized compression, 210 (7 sessions of 30, with breaks in between) pairs of sound files consisting of the spoken sentences, discussed earlier, in noisy (babble) background were played at SNR of 0 dB. The participants were asked to indicate which sound file or clip they preferred or whether both sound clips sounded the same to them. It is worth mentioning here that increasing the number of sound files naturally improves the training and 210 audio sound files may not be adequate to cover all possible combinations of compression settings. For example, for $\beta = 2$, the number of possible actions is $2^5 = 32$, demanding $\binom{32}{2} = 992$ pairs of sound files. However, to avoid human fatigue, the test sessions were limited to 2 hours. This time frame constrained the number of audio sound files to 210 over 7 sessions. The data augmentation mentioned earlier was then applied to the collected dataset $D$ of triplets $(c^1, c^2, \mu)$. The reward predictor was trained based on the augmented data to learn a participant's hearing preferences.

After training the policy, a comparison test was conducted between the personalized compression and the DSL-v5 reference prescriptive compression by playing 60 randomly selected sentences across different talkers in a noisy (babble) background at the same SNR level of 0dB. Table III provides the compression ratios of DSL-v5 versus the compression rations of the developed personalized approach for five participants with mild to moderate bilateral hearing loss who took part in this study. The outcomes of the participant testing experiments in terms of preference percentages are shown as a bar chart in Fig. 10. As can be seen from this figure, on average, personalized settings were clearly preferred by the participants over the DSL-v5 settings across different talkers and sentences heard. In other words, the number of times the personalized settings were preferred by the participants were nearly 7 times greater than the number of times the DSL-v5 settings were preferred. These results indicate that the developed personalized or individualized compression indeed is more effective than a one-size-fits-all DSL-v5 prescriptive compression approach.

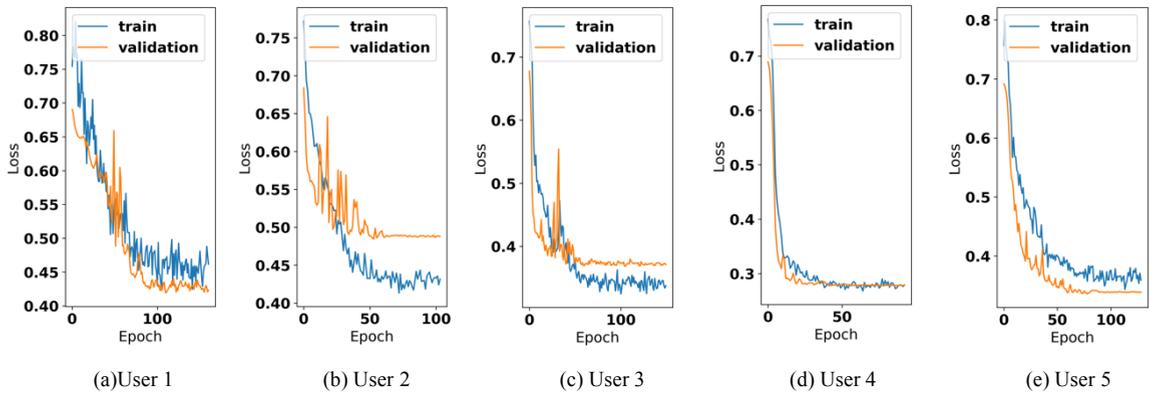

FIGURE 7. Cross-entropy loss value in training reward predictor for simulated (a) user 1, (b) user 2, (c) user 3, (d) user 4, and (e) user 5.

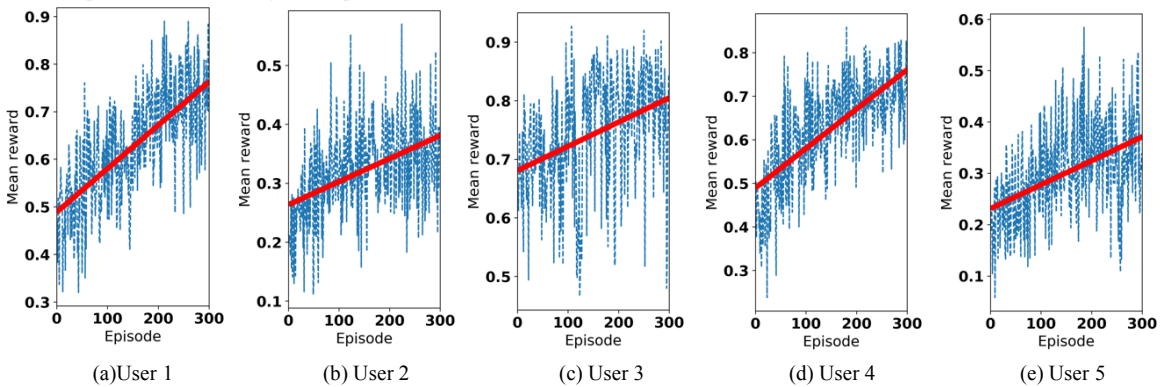

FIGURE 8. Mean of normalized reward per episode and its trend (in solid red line) in the agent training for (a) simulated user 1, (b) simulated user 2, (c) simulated user 3, (d) simulated user 4, and (e) simulated user 5.

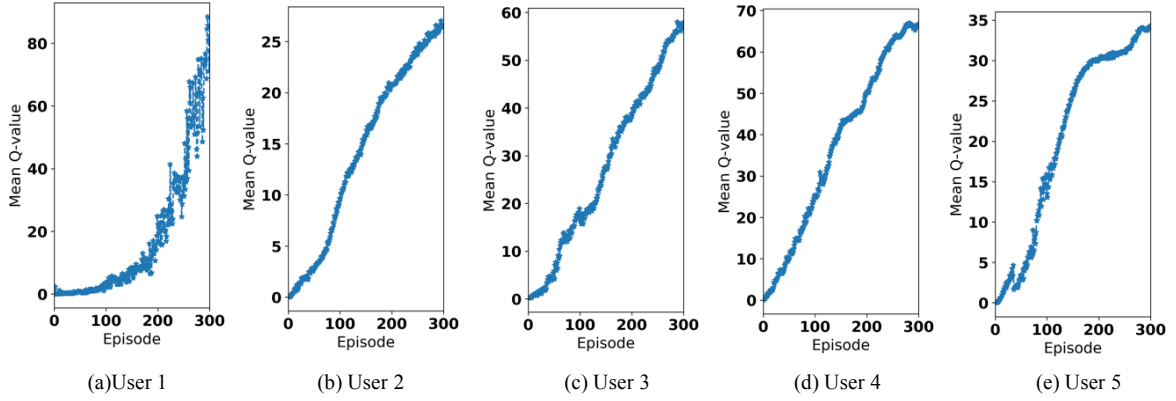

(a)User 1  (b) User 2  (c) User 3  (d) User 4  (e) User 5

**FIGURE 9.** Mean Q-value per episode in the agent training for (a) simulated user 1, (b) simulated user 2, (c) simulated user 3, (d) simulated user 4, and (e) simulated user 5.

TABLE III
SUBJECT TESTING EXPERIMENTS: DSL-V5 VS. PERSONALIZED COMPRESSION RATIOS.

| Subject | Level of hearing loss | Audiogram in freq. bands [0.5 1.0 2.0 4.0 6.0] kHz | DSL-v5 gains for soft speech | DSL-v5 compression ratios | Personalized compression ratios |
|---|---|---|---|---|---|
| 1 | Mild | [15, 20, 20, 30, 30] | [7, 8, 14, 17, 15] | [1.1, 1.2, 1.3, 1.2, 1.3] | [1.1, 1.2, 1.3, **4.8**, **5.2**] |
| 2 | Mild | [15, 15, 20, 20, 30] | [5, 6, 14, 15, 15] | [1.1, 1.2, 1.3, 1.2, 1.2] | [**4.4**, 1.2, **5.2**, 1.2, **4.8**] |
| 3 | Moderate | [20, 20, 40, 50, 60] | [11, 12, 24, 29, 34] | [1.1, 1.2, 1.3, 1.2, 1.4] | [**4.4**, 1.2, 1.3, **4.8**, **5.6**] |
| 4 | Mild | [25, 20, 20, 40, 30] | [13, 11, 14, 22, 15] | [1.1, 1.3, 1.3, 1.3, 1.3] | [**4.4**, 1.3, 1.3, **5.2**, 1.3] |
| 5 | Moderate | [20, 20, 30, 40, 40] | [6, 11, 20, 23, 20] | [1.1, 1.2, 1.3, 1.2, 1.4] | [1.1, 1.2, 1.3, **4.8**, **5.6**] |

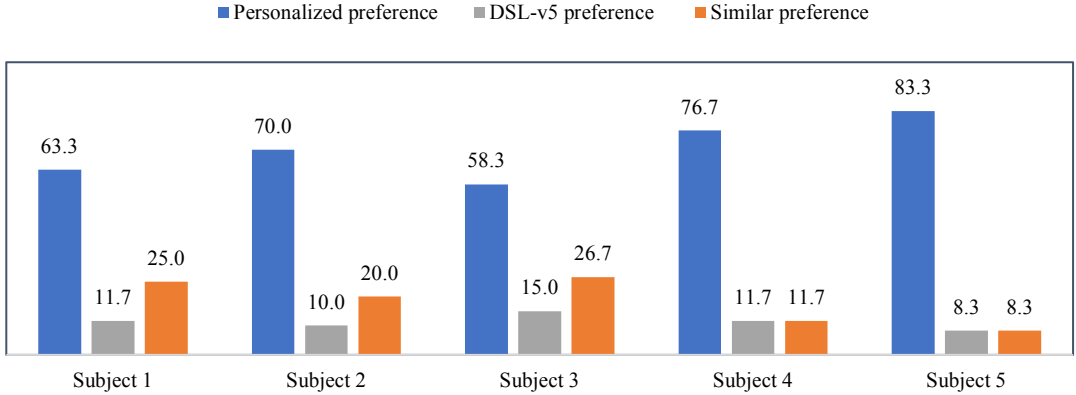

**FIGURE 10.** Outcome of subject testing experiments in percentages: comparison of hearing preference between personalized compression and DSL-v5 compression.

## IV. CONCLUSION

In this paper, an active human-in-loop deep reinforcement learning-based personalized hearing aid fitting approach is developed to improve the currently practiced one-size-fits-all hearing aid fitting. The current fitting practice involves setting compression gains based on gain averages of a group of users which are not necessarily optimal for a specific user. The developed approach personalizes fitting by active learning via a deep reinforcement learning framework. This is the first time deep and reinforcement learning has been used to achieve improved hearing aid compression. Both simulation and experimental results show the effectiveness of the developed personalized approach in achieving preferred hearing outcomes. In our future work, it is planned to extend the developed personalization approach to the other parameters of compression.

## ACKNOWLEDGMENT

We wish to express our appreciation to Ms. Tina Campbell and Dr. Celia Escabi for their help in putting together the IRB materials and their assistance with the subject testing. We also thank Mr. Ali Salman for his help with the codes written for this project.